# Selective laser ablation of metal thin films
# using ultrashort pulses

*Byunggi Kim[a], Han Ku Nam[a], Shotaro Watanabe[b], Sanguk Park[a], Yunseok Kim[c], Young-Jin Kim[a],*

*Kazuyoshi Fushinobu[b], \*Seung-Woo Kim[a]*

[a]*Department of Mechanical Engineering, Korea Advanced Institute of Science and Technology, 291 Daehak-ro, Yuseong-gu, Daejeon, 34141, Republic of Korea*

[b]*Department of Mechanical Engineering, Tokyo Institute of Technology, 2-12-1 Ookayama, Meguro-ku 152-8550, Japan*

[c]*Lasernics Co., KAIST ICC, 193 Munji-ro, Yuseong-gu, Daejeon, Republic of Korea*

*\*Corresponding author: swk@kaist.ac.kr (S.-W. Kim)*

## ABSTRACT

**Selective thin-film removal is needed in many microfabrication processes such as 3-D patterning of optoelectronic devices and localized repairing of integrated circuits. Various wet or dry etching methods are available, but laser machining is a tool of green manufacturing as it can remove thin films by ablation without use of toxic chemicals. However, laser ablation causes thermal damage on neighboring patterns and underneath substrates, hindering its extensive use with high precision and integrity. Here, using ultrashort laser pulses of sub-picosecond duration, we demonstrate an ultrafast mechanism of laser ablation that leads to selective removal of a thin metal film with minimal damage on the substrate. The ultrafast laser ablation is accomplished with the insertion of a transition metal interlayer that offers high electron-phonon coupling to trigger vaporization in a picosecond timescale. This contained form of heat transfer permits lifting off the metal thin-film layer while blocking heat conduction to the substrate. Our ultrafast scheme of selective thin film removal is analytically validated using a two-temperature model of heat transfer between electrons and phonons in material. Further, experimental verification is made using 0.2 ps laser pulses by micropatterning metal films for various applications.**

## 1. Introduction

Laser ablation is preferably used for clean machining of material without use of toxic chemicals, but special care is required to minimize thermal damage caused by excessive heat input. Ultrashort lasers of sub-picosecond pulse duration are now emerging as an enabling tool of micromachining with a potential of non-thermal ablation by ultrafast control of pulse energy and timing [1–3]. In view of microscopic heat transfer, the intense ultrashort light pulse is instantly absorbed by electrons in a picosecond timescale, provoking ultrafast electron-to-electron heat transfer before a thermal equilibrium is reached between electrons and lattices of material. The ultrafast energy absorption by photoelectrons consequently reduces the heat affected zone by ablating material with extreme heat concentration ahead of thermal diffusion. This concept of clean laser ablation has been superbly demonstrated by expeditious gigahertz bursting of femtosecond light pulses [1]. Besides, by chirped pulse amplification, ultrashort lasers can readily offer very high pulse intensities to stimulate nonlinear absorption by means of multiphoton or tunneling effects. The consequence is a possibility of non-thermal clean ablation in the form of electrons ionization and coulomb explosion particularly for dielectric and semiconductor materials [4, 5].

Among diverse demands on clean laser ablation, micropatterning on thin metal films is a special area where nanosecond pulse lasers have long been used with a capability of fluence control to reduce the heat affected zone [6]. Ultrashort lasers recently began to be adopted to accomplish more precise fluence control by ultrafast regulation of the pulse energy and repetition rate [7–9]. Meanwhile, several intriguing behaviors of ultrafast photoelectrons have been uncovered for better understanding of the heat transfer mechanism occurring within thin film metals: First, compared to the case of using continuous-wave or nanosecond lasers, ultrashort pulses can penetrate much faster, deeper towards the bottom of a metal film. This phenomenon is explained as a result of ballistic transport of photoelectrons excited so drastically to a thermally non-equilibrium state with lattices [10, 11]. Second, the temperature development within a thin metal film irradiated by ultrashort lasers alters significantly with the insertion of a transition metal interlayer that provides strong electron-phonon coupling [12]. The interlayer consequently heats up more quickly than the top metal film to yield a reversed temperature distribution. Third, the metal-dielectric interface weakens both the electron-electron and electron-phonon couplings, while the interfacial heat flux is governed by direct scattering of excited electrons from metal to dielectric along with resistive phonon-phonon coupling [13–15]. Additionally, ultrafast clean ablation draws much attention to benefit the nano-scale repair of thin metal films within integrated circuits as the task has to be fulfilled with as less damage as possible on underneath patterns or substrates [16]

In this study, based on the above findings of extraordinary behaviors of ultrafast photoelectrons, we propose a noble mechanism to remove thin metal films with minimal damage on the underneath substrate. This selective metal film removal mechanism is validated by numerical simulation based on a two-temperature model of heat transfer between electrons and phonons. In parallel, experimental verification is made using 0.2 ps laser pulses on a 100 nm thick gold ($Au$) film deposited on a glass substrate with a 5 nm titanium ($Ti$) interlayer. This investigation confirms that the mechanism of selective metal film ablation is facilitated with the insertion of a transition metal interlayer of strong electron-phonon coupling that eventually renders the thin-film layer be lifted off by instantaneous vaporization, thereby blocking thermal conduction to the substrate. Lastly, a series of clean ablation examples are fabricated to prove that ultrafast laser ablation can be used for clean micropatterning on diverse metal films without use of noxious wet or dry etching.

## 2. Heat Transfer Analysis

Figure **1a** illustrates thermal conduction paths on a target specimen considered in this study to investigate the mechanism of selective thin film removal with the injection of ultrashort light pulses. The target specimen consists of a metal film deposited on a dielectric substrate with an adhesion interlayer in the middle. The incident photon energy is absorbed by free electrons in the metal film and assumed to spread through three distinct modes of quantum coupling; the electron-electron (**e-e**) coupling, electron-phonon (**e-p**) coupling and phonon-phonon (**p-p**) coupling. The electron temperature raised upon absorption of a laser pulse is redistributed to neighboring electrons by the **e-e** coupling and to lattices by the **e-p** coupling. The lattice temperature rises in proportion to the **e-p** coupling and is transferred to other lattices by the **p-p** coupling with heat dissipation over the entire metal layer. In the dielectric substrate, the **e-e** coupling stops due to the absence of free electrons, while the **e-p** coupling weakens but continues by direct scattering of hot electrons from the metal-dielectric interface. Meanwhile, the **p-p** coupling transmits through the metal-dielectric interface with some resistance.

The top metal film of the target specimen is made of a noble metal, such as gold (*Au*), silver (*Ag*) and copper (*Cu*). In solid-state physics, noble metals are known to provide high thermal and electric conductivity with their *s/p*-bands being positioned near the Fermi level, while the *d*-band of high density of states is placed far below the Fermi level. In contrast, the interlayer underneath the top metal film is built with a transition metal, such as titanium (*Ti*) or chrome (*Cr*), acting as an adhesion layer offering strong material bonding and heat adsorption. Transition metals usually show their partially-filled *d*-band near the Fermi level and exhibit strong **e-p** coupling to allow for fast thermal relaxation of excited electrons to lattices as explained in the *d*-band theory [17]. To be specific, *Ti* has a strong **e-p** coupling coefficient of two orders of magnitude larger than *Au* [18, 19]. This implies that the heat of excited *Ti* electrons rapidly transmits to *Ti* lattices. Consequently, the *Ti* lattice temperature rises more quickly than the *Au* lattice temperature, reaching the vaporizing ablation point to separate the metal film layer from the substrate.

Figure **1b** shows a computational result of the lattice temperature at the very instant of film ablation, which is plotted along the depth direction starting from the top surface of an *Au* film (100 nm thickness) through a *Ti* interlayer (5 nm thickness) towards the substrate structure of aluminosilicate glass (200 nm thickness). The incident pulse laser offers a constant fluence of 3.9 J/cm$^2$ with a center wavelength of 1035 nm. The lattice temperature distribution was calculated for two different cases of pulse duration; one is as short as 0.2 ps (red curve) and the other is 100 ps (blue curve). The computation was performed using the two-temperature model configured to deal with the **e-e** coupling, **e-p** coupling and **p-p** coupling all together as described in **Appendix**. The two-temperature model handles the electron temperature separately from the lattice temperature, enabling the role of the interlayer more precisely in the non-equilibrium thermal transfer between the metal film and the dielectric substrate. The result confirms that for the shorter pulse of sub-

picosecond duration, the *Ti* lattice temperature surpasses the vaporization temperature before the *Au* metal film reaches its melting temperature. This reversed temperature development makes the *Au* film be selectively lifted off by the vaporization of the *Ti* interlayer. Note that the dielectric substrate is not subject to a substantial temperature rise as not much time is allowed for heat transfer. In comparison, for the longer pulse of 100 ps duration, the *Ti* lattice temperature grows at a moderate pace comparable to the *Au* lattice temperature, being not able to reach the vaporization temperature. In consequence, the *Au* film is ablated together with the *Ti* interlayer by melting almost at a same instance. The slow ablation allows a large amount of heat to infiltrate the substrate with succeeding temperature rise above the glass transition point leading to thermal damage.

Figure **1c** presents the timescale of film removal estimated using the two-temperature model of heat transfer described in **Appendix**. For the short pulse of 0.2 ps duration, the **e-e** coupling initiates immediately upon irradiation of the laser pulse and continues to propagate throughout the *Au* film and the *Ti* interlayer during a few tens of picoseconds after the laser pulse retreats. The **e-e** coupling takes place with a high electron-to-electron thermal conductivity, being further accelerated by the ballistic transport of photoelectrons excited drastically to a thermally non-equilibrium state with lattices [10, 11]. The fast **e-e** coupling is then accompanied by the **e-p** coupling that raises the lattice temperature. With a higher **e-p** coupling coefficient, the *Ti* lattice temperature begins to rise ahead of the *Au* lattice temperature, reaching the vaporizing point. Even after the lift-off removal of the *Au* film, the lattice temperature in the surrounding *Au* film remains high with continuation of the **e-p** coupling, forming a crater later by cooling. In comparison, for the long pulse of 100 ps duration, both the **e-e** and **e-p** couplings progress slowly all the way along the whole duration of the laser pulse. The lattice temperature rise is consequently slowed and not able to reach the vaporizing point. The *Au* film and *Ti* interlayer are ablated by melting long after the laser pulse ends with a long-elapsed time of ∼ 1,000 ps, with a substantial amount of heat transmitted into the dielectric substrate.

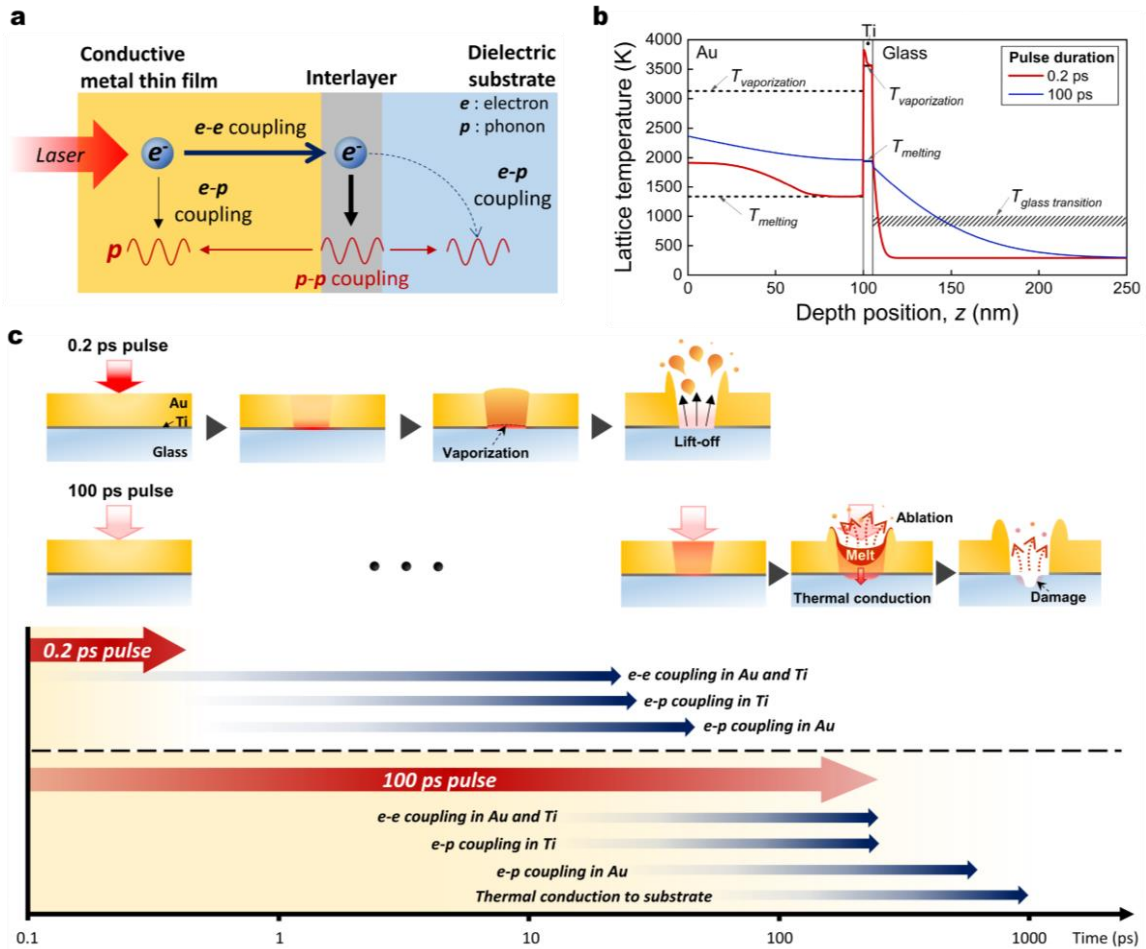

**(2 column) Figure 1 |** Selective thin film removal by laser ablation using ultrashort pulses. **a,** Heat transfer paths inside the target specimen comprised of a thin metal film deposited on a dielectric substrate with an interlayer in between. **b** Lattice temperature distribution calculated along the depth position using a two-temperature model of heat transfer for solids. The target specimen consists a 100 nm gold (*Au*) film, a 10 nm titanium (*Ti*) interlayer, and a glass substrate. The incident laser offers a fluence of 3.9 J/cm² with an ultrashort pulse duration of 0.2 ps, leading to a drastic temperature rise in the *Ti* interlayer. Another temperature profile (blue) calculated for a stretched pulse of 100 ps duration shows no temperature rise. **c** Heat transfer timescales revealing ultrafast selective thin film removal only for ultrashort laser pulses.

## 3. Results and Discussion

Figure **2** shows the experimental data obtained with an apparatus configured in Fig. **2a** to validate the mechanism of selective thin-film removal proposed in this study. The light source is a home-made Yb-doped fiber laser emitting 0.2 ps pulses at a center wavelength of 1035 nm at a repetition rate of 1 MHz. The pulse energy is tunable to a maximum of 10 µJ per pulse through chirped pulse amplification (CPA). An electro-optic modulator (EOM) is employed for fast on-off pulse picking control by polarization rotation, so an event of single-pulse ablation at a time can be made using a high-speed arbitrary waveform generator (AWG). The pulse duration can be varied from 0.2 ps to 84.6 ps using an adjustable compression controller that is equipped with two gratings and a roof mirror on a translation linear stage. The output laser beam is focused on a 4.6 µm spot diameter through an achromatic objective of 5X magnification. To be consistent with the simulation conditions of Fig. **1b**, the specimen shown in Fig. **2b** is prepared with an *Au* film of 100 nm thickness, a *Ti* interlayer of 5 nm thickness, and an aluminosilicate glass substrate. The specimen is mounted on a five-axis stage to conduct precise positioning control during laser exposure with the aid of a digital camera under illumination of a halogen lamp.

The crater depth created by a single-shot pulse ablation was measured using a confocal optical microscope. The crater depth was defined at the deepest spot of the crater with respect to the undamaged flat top surface. Then the measured depths were plotted in terms of the pulse duration for various laser fluences as in Fig. **2c**. The plotted result verifies that the intended mechanism of selective *Au* film removal is achieved when the pulse duration is taken shorter than a few picoseconds under well-adjusted fluence control in the range of 3.2 ~ 3.9 J/cm². For lower fluences, the *Au* film is not completely removed as the incident energy is not enough. When the pulse duration is taken longer than a few picoseconds, no selective ablation is observed at all as the glass substrate suffers excessive removal even for moderate laser fluences. More specifically, as revealed in detailed 3-D profiles of Fig. **2d** measured using an atomic force microscope, the craters exhibit a flat bottom profile without significant substrate damage when the pulse duration is kept in the range of 0.2 – 6.35 ps with optimized fluence of 3.9 J/cm². In contrast, for a longer pulse duration of 84.6 ps, the glass substrate becomes overcut even for the same fluence. Furthermore, for a higher fluence of 5.2 J/cm² as illustrated in Fig. **2e**, the substrate damage is unavoidable even the pulse duration is taken as short as 0.2 ps. This implies that excessive fluence coupled with very short pulse duration accelerates the direct scattering of hot electrons into the substrate, creating ejecta such as plume or particles making the crater bottom profile be more irregular.

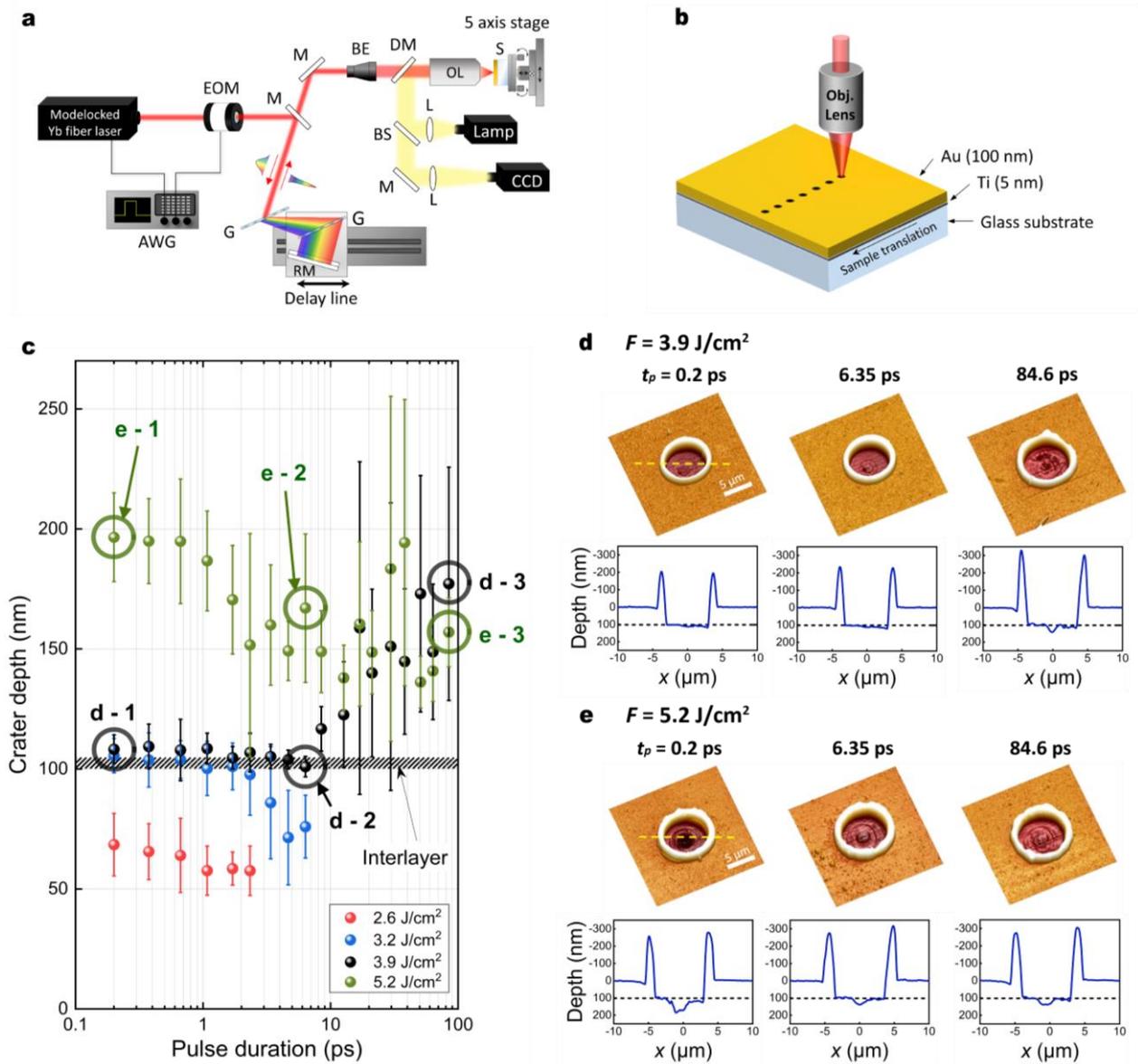

**(2 column) Figure 2** | Single pulse ablation experiment for selective thin film removal. **a** Apparatus setup with the tunable compression capability of the pulse duration from 0.2 ps to 100 ps. AWG: arbitrary waveform generator, EOM: electro-optic modulator, M: mirror, G: grating, RM: roof mirror, BE: beam expander, DM: dichroic mirror, BS: beam splitter, L: lens, OL: objective lens, S: specimen. **b** Single pulse irradiation on *Au-Ti*-glass specimens with varying the pulse duration and fluence. **c** Experiment results plotted in terms of the crater depth vs. pulse duration. **d** & **e** Crater profiles measured by atomic force microscopy (AFM) for an optimum fluence of 3.9 J/cm² in comparison with a higher one of 5.2 J/cm². Horizontal dashed lines indicate the film thickness 0f 100 nm. *F* and $t_p$ are the fluence and pulse duration, respectively.

Figure **3** depicts the *Ti* interlayer's role in the ablation mechanism of selective thin film removal as it is further detailed by simulation using the two-temperature model of heat transfer given in Appendix. The thickness effect of the *Ti* interlayer on the overall lattice temperature distribution was calculated for two comparative cases of pulse duration; 0.2 ps in Fig. **3a** and 84.6 ps in Fig. **3b**. Laser fluence was fixed at 3.9 J/cm$^2$ for both the cases. The former result implicates that the amount of heat absorbed by the *Ti* interlayer grows with increasing the interlayer thickness, but the lattice temperature of the glass substrate is not affected significantly and maintained below the transition temperature. On the other hand, the latter case of 84.6 ps pulse reveals that the glass temperature exceeds the transition temperature, confirming the experimental observation results of Fig. **2c-e**.

Fig. **3c** shows the temporal evolution of the electron temperature of the *Ti* interlayer of a thickness of 5 nm upon irradiation of a 0.2 ps, 3.9 J/cm$^2$ pulse. Immediately after the laser is turned off, the *Ti* interlayer temperature increases drastically by the thermal conduction from the *Au* electrons. The electron temperature of *Ti* increases again with a time delay of 20 ps due to the latent heat consumed by the *Au* layer during its phase change of melting. With strong **e-p** coupling, the calculated result indicates that the *Ti* lattice temperature catches up with the electron temperature to reach a thermal equilibrium state within 30 ps. In addition to titanium (*Ti*), chromium (*Cr*) is also widely used as an adhesion interlayer material but its **e-p** coupling is not as strong as that of *Ti*. Thus, for a same incident laser fluence, the *Cr* electron temperature rises higher than that of *Ti* while the *Cr* lattice temperature stays below the vaporization temperature in a thermal non-equilibrium state (Fig. **3d**). This situation makes the lift-off removal by vaporization more difficult. Further, the high electron temperature accelerates direct scattering of hot electrons from the Cr interlayer to the dielectric substrate, causing more severe thermal damage.

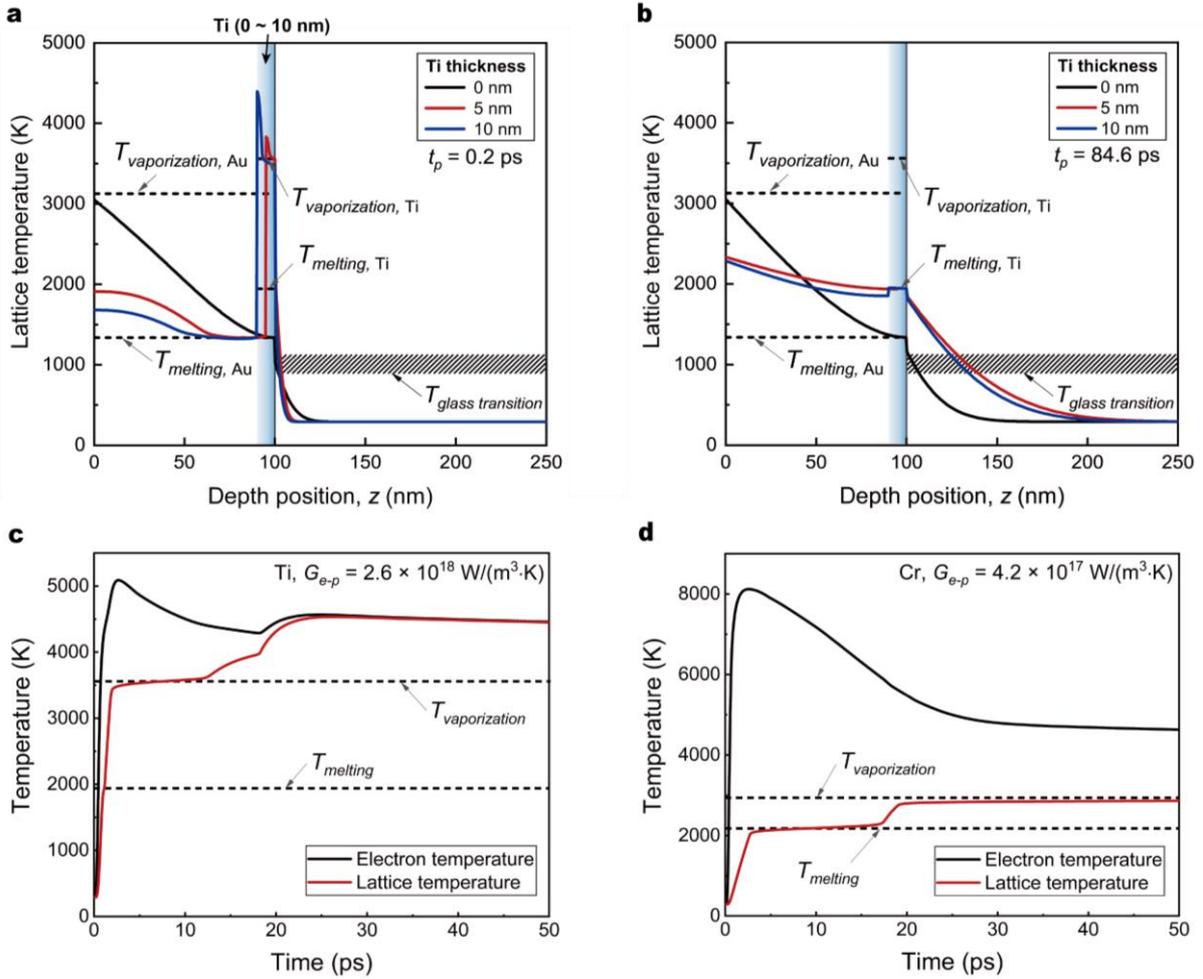

**(2 column) Figure 3** | Interlayer effect on selective thin film removal. **a** & **b** Lattice temperature variation with *Ti* interlayer thickness. The pulse duration $t_p$ is 0.2 ps (left) or 84.6 ps (right). Laser fluence is fixed at 3.9 J/cm² for both. **c** & **d** Temporal evolution of electron temperature and lattice temperature for *Ti* interlayer (left) and *Cr* interlayer (right). The pulse duration is 0.2 ps for both. $G_{e-p}$: electron-phonon coupling coefficient, T: temperature.

Figure **4** shows a series of clean ablation examples fabricated to validate the ultrafast mechanism of selective laser ablation, of which the machining conditions are summarized in Table 1. First, microdots of 4.6 μm diameter made by single-shot ablation sequentially on a 100 nm *Au* thin film (Fig. **4a**) display no thermal damage on the glass substrate as confirmed by optical and scanning electron microscope (SEM, inset) images. Second, line patterns of a 2 μm width formed by ultrafast selective ablation (Fig. **4b**) are compared with other line patterns made by backside scribing (Fig. **4c**). Due to the difficulty achieving clean ablation using nanosecond lasers, the backside scribing has long been investigated by injecting the laser beam backwards from the glass substrate so as to ablate the metal film layer mechanically by raising the temperature at the vicinity of the film-substrate interface [6, 20]. The backside scribing permits strong thermal stress to be induced with less laser pulse energy compared to the front scribing, thereby reducing thermal damage on the substrate. However, the backside scribing can be used only for rigid, transparent substrates, while the ultrashort selective ablation proposed in this study is applicable without the restriction, producing more even surface morphology as verified in SEM images.

Many practical applications of selective clean ablation can be found for electronic industries, particularly for the nano-scale repair of thin film copper layers in integrated circuits (Fig. **4d**). Without notable thermal damage on the silicon nitride sublayer, top copper layers can be cut off with a sub-μm linewidth as verified in optical and SEM images. Periodic gratings fabricated on a large area by laser beam scanning (Fig. **4e**) verify that *Au* thin films can be removed with well-defined edge shapes with no damage noticeable on the glass substrate. The Raman spectrum obtained from the glass substrate (Fig. **4f**) after scribing with a 0.2 ps laser indicates no difference from the pure glass substrate, whereas the same scribing with a 4 ns laser suffers significant material modification due to excessive heat. The optical transmission testing (Fig. **4g**) also reveals no deterioration for the case of 0.2 ps laser scribing. Finally, a honeycomb-shaped *Au* metal grid patterned on a flexible polyethylene terephthalate (PET) substrate (Fig. **4h**) maintains the original flexibility as well as the light transmission capability with minimal substrate damage only near the foot of the metal grid.

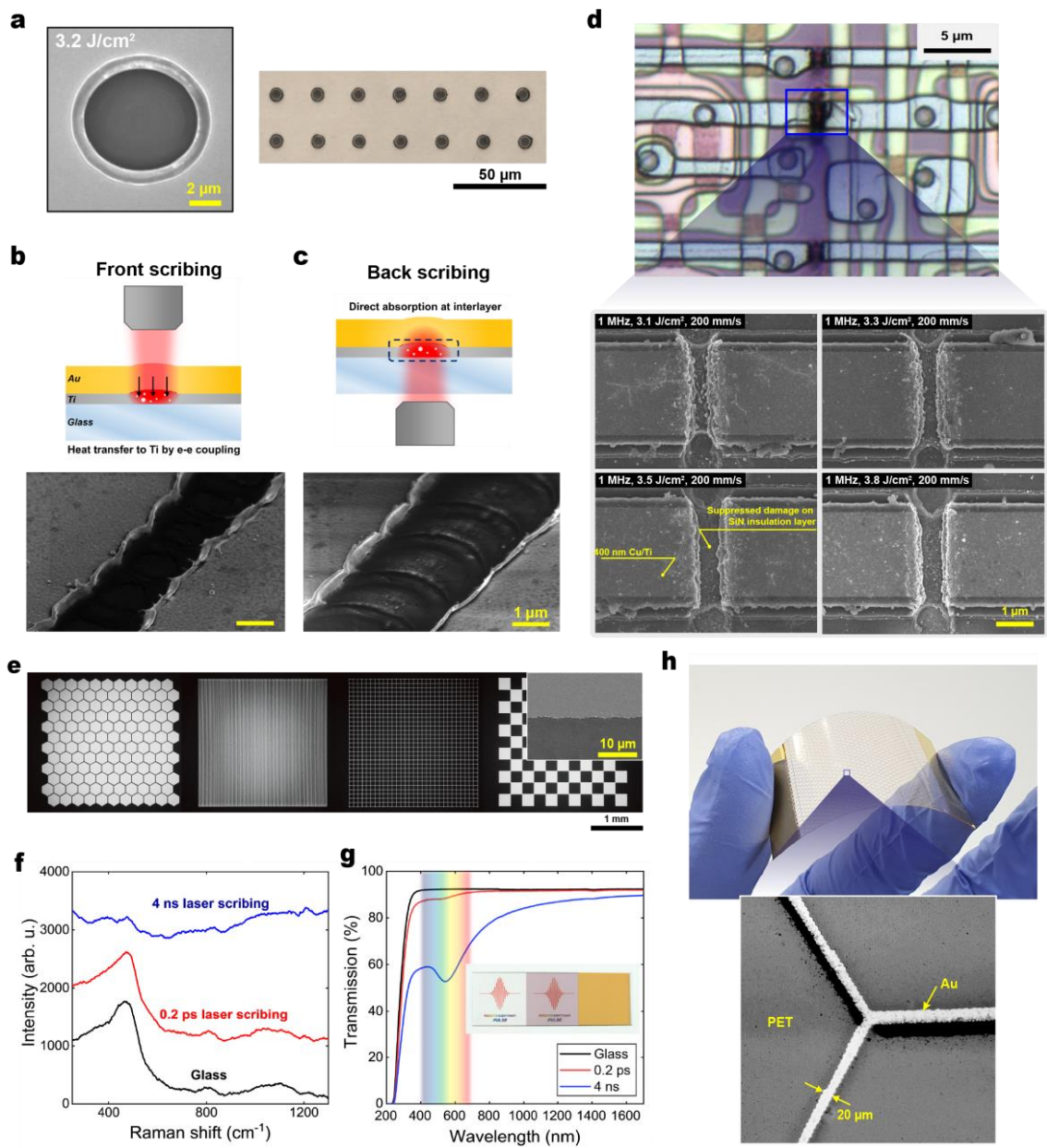

**(2 column) Figure 4** | Machining examples of selective thin film removal. **a** Single-shot microdots array made on a 100 nm gold (*Au*) thin film. Inset shows a scanning electron microscope (SEM) image. **b** & **c** Frontside and backside scribing. **d** Repair of integrated circuit patterns. **e** Periodic patterns made on a 50 nm gold thin film. **f** & **g** Glass substrate damage characterization using Raman spectroscopy and transmission testing. **h** Honeycomb-shaped *Au* metal grid patterned on a polyethylene terephthalate (PET) flexible substrate.

**Table 1**. Machining conditions for applications

| Machining example | Figure | Thin film dimensions | Laser pulse duration | Laser wavelength | Laser pulse repetition rate | Beam diameter ($1/e^2$) | Laser fluence | Scanning speed | Hatch |
|---|---|---|---|---|---|---|---|---|---|
| Microdots | Fig. 4a | 100 nm *Au*, 5 nm *Ti* on glass | 0.2 ps | 1035 nm | - | 4.6 μm | 3.2 J/cm² | - | - |
| Line scribing | Fig. 4b, 4c | 50 nm *Au*, 2.5 nm *Ti* on glass | 0.2 ps | 1040 nm | 200 kHz | 1.4 μm | 2.5 J/cm² | 200 mm/s | - |
| Repair of integrated circuits | Fig. 4d | 400 nm *Cu/Ti* interlayer on *SiN* | 0.2 ps | 1040 nm | 1 MHz | 1.4 μm | 2.5 J/cm² | 200 mm/s | - |
| Large area patterns | Fig. 4e | 50 nm *Au*, 5 nm *Ti* on glass | 0.2 ps | 1040 nm | 200 kHz | 12 μm | 3.6 J/cm² | 1250 mm/s | 6 μm |
| | | | 4 ns | 1064 nm | 200 kHz | 30 μm | 9.9 J/cm² | 1250 mm/s | 20 μm |
| Honeycomb metal grid | Fig. 4h | 100 nm *Au*, 25 nm *ITO* on *PET* | 0.2 ps | 1040 nm | 200 kHz | 12 μm | 2.2 J/cm² | 1250 mm/s | 7 μm |

## 4. Conclusions

We have investigated analytically and experimentally how ultrashort lasers permit selective removal of thin metal films with minimal substrate damage. Computational simulation performed using a two-temperature model of heat transfer reveals that ultrashort pulses of 0.2 ps duration induce a thermal non-equilibrium between electrons and lattices in the *Au* metal film, boosting heat transfer to the *Ti* interlayer with drastic temperature rise by strong electron-phonon (**e-p**) coupling. The *Ti* lattice temperature surpasses the vaporization temperature to remove the *Au* film by lift off, thereby blocking the heat conduction to the glass substrate within a few tens of picoseconds. Experimental verification confirms the ultrafast mechanism of selective metal film ablation facilitated with the insertion of a transition metal interlayer of strong electron-phonon coupling. Lastly, a series of clean ablation examples proves that ultrafast laser ablation can be used for clean micropatterning on diverse metal films including the repair of integrated circuits. Our study conducted using a simple two-layer configuration is expected to be applicable to more complicated multilayer structures widely used for high-quality, high-speed processing of optoelectronic devices.

## Appendix. Numerical methods

Temperature development in the interlayer in the ultrafast regime is strongly influenced by the **e-e** coupling as well as the **e-p** coupling in a strong thermal non-equilibrium state between electrons and lattices. Therefore, a two-temperature model needs to be adopted to account for the heat transport of electrons and lattices separately. The Boltzmann transport theory [21] may be also used for the two-temperature analysis by means of statistical dynamics, but the Fourier heat theory is more preferable as it permits calculating macroscopic temperature distribution leading to ablation with less computational burden. In heat transfer by conduction within a solid, the electron temperature $(T_e)$ and the lattice temperature $(T_l)$ are described with two coupled equations of thermal diffusion [22]:

$$C_e \frac{\partial T_e}{\partial t} = \nabla(k_e \nabla T_e) - G_{e-p} \times (T_e - T_l) + S \qquad (A.1)$$

$$\rho_l \left[ c_{pl} + L_m \delta_m + L_v \delta_v \right] \frac{\partial T_l}{\partial t} = \nabla(k_l \nabla T_l) + G_{e-p} \times (T_e - T_l) \qquad (A.2)$$

The subscripts $e$ and $l$ denote the electron and lattice, respectively. The parameters $C_e$, $k$, $\rho$ and $c_p$ are the free electron specific heat, the thermal conductivity, the material density, and the specific heat. In addition, $L_m \delta_m$ represents the latent heat for melting and $L_v \delta_v$ the latent heat for vaporization, with the Kronecker $\delta$-like function being defined as

$$\delta(T_l - T_i, \Delta) = \frac{1}{\sqrt{2\pi}\Delta} exp \left[ -\frac{(T_l - T_i)^2}{2\Delta^2} \right] \quad . \qquad (A.3)$$

$T_i$ represents either the meting temperature or the vaporization temperature, where $\Delta$ denotes the full-width-at-half-maximum (FWHM) temperature range over which the phase change of melting or vaporization occurs.

The source term $S$ of Eq. (A.1) has units of W/m$^3$ and is given as a function of the depth $z$ measured from the top metal surface towards the substrate;

$$S = \frac{1-R}{1 - \exp\left[-\frac{L}{(d+d_b)}\right]} \times 4 \times \sqrt{\frac{4\ln 2}{\pi}} \times \frac{F}{(d+d_b)t_p} \times$$

$$\exp\left[ -\frac{z}{d+d_b} - 4\ln 2\left(\frac{t-2t_p}{t_p}\right)^2 \right] \qquad (A.4)$$

The parameters $R$, $L$, $F$ and $t_p$ are the reflection coefficient, the film thickness, the laser incident fluence, and the laser pulse duration, respectively. The parameter $d$ is the optical penetration depth calculated by linear light propagation theory. The other parameter $d_b$ indicates the mean free path of excited free electrons extended additionally by ultrashort intense light pulses, which is called the ballistic range of electrons. In this study, $d_b$ was estimated to be ~ 100 nm for *Au* [10].

At a metal-metal interface, a certain amount of thermal resistance needs to be considered in dealing with electron–electron coupling. In this study, the electron thermal resistance $R_{ee}$ between the *Au* film and the *Ti* interlayer was determined using the extended diffusive mismatch model of electron heat transfer at a metal–metal interface [12, 23] as follows:

$$R_{ee,Au/Ti} = \frac{4(Z_{Au} + Z_{Ti})}{Z_{Au} Z_{Ti}} \qquad (A.5)$$

where $Z = C_e v_e$ with $v_e$ being the electron velocity usually close to the Fermi velocity. Between the *Ti* interlayer and the glass substrate, $R_{ee}$ was assumed infinite since the latter is dielectric with no free electrons.

both the metal–dielectric and metal–metal interfaces, the phonon thermal resistance $R_{pp}$ is found to have little influence on the temperature distribution in the ultrafast regime [13]. In this study, $R_{pp}$ is taken as $4 \times 10^{-9}$ m$^2$ K/W with reference to the literature data obtained using pump-probe spectroscopy [15] and quantum mechanical calculation [12].

The temperature-dependent effect on the free-electron heat capacity was considered as $C_e = \gamma T_e$ with $\gamma$ being assumed as a constant. The electron–phonon coupling coefficient $G_{e\text{-}p}$ was also modified as $G_{e-p} = G_0 \left[ \frac{A_e}{B_l} (T_e + T_l) + 1 \right]$ with $G_0$ being the room-temperature value of $G_{e\text{-}p}$. And $A_e = 1.2 \times 10^7$ K$^{-2}$s$^{-1}$ and $B_l = 1.2 \times 10^{11}$ K$^{-1}$s$^{-1}$ are the material-dependent constants for the *Au* electron and lattice, respectively [24, 25]. The thermal conductivity of electron was determined as [26]:

$$k_e = \chi \frac{(\vartheta_e^2 + 0.16)^{5/4} (\vartheta_e^2 + 0.44) \vartheta_e}{\sqrt{\vartheta_e^2 + 0.092} (\vartheta_e^2 + \eta \vartheta_l)} \qquad (A.6)$$

Here, $\vartheta_e = T_e/T_F$ and $\vartheta_l = T_l/T_F$ are the dimensionless electron and lattice temperatures normalized with regards to the Fermi temperature $T_F$. The fitting parameters are $\chi = 353$ W/(m·K) and $\eta = 0.16$ for *Au*. Along with Eqs. (A.5) and (A.6), the temperature dependence of $G_{e\text{-}p}$ was considered only for *Au* while $G_{e\text{-}p}$ was assumed constant for *Ti* and *Cr* due to the lack of available data. The thermal conductivity of electrons for *Ti* and *Cr* was assumed the same as that of the lattice [18,19]. Table A1 lists the thermophysical properties of the materials used in this study.

The film thickness dealt with in this study is much smaller than the laser beam diameter. Thus, with irradiation of a single laser pulse, Eqs (A.1) and (A.2) are solved numerically by one-dimensional finite difference calculation along the depth z-direction. Implicit numerical computation was made with a relative residual of $10^{-7}$. Direct electrons scattering from metal to dielectric may be significant, particularly for ultrashort pulses yielding high electron temperatures. However, the electron-phonon coupling of the interlayer is assumed so strong as to neglect direct electrons scattering from the interlayer to the dielectric substrate. Natural convection, thermal conduction to ambient air, and radiation heat transfer are neglected. Only the thermal conduction from the top metal film through the interlayer to the substrate are considered with zero heat flux boundary conditions;

$$\frac{\partial T_e}{\partial z}\Big|_{flim\ surface} = \frac{\partial T_l}{\partial z}\Big|_{film\ surface} = \frac{\partial T_g}{\partial z}\Big|_{z=\infty} = 0 \tag{A7}$$

**Table A1**. Thermophysical properties of the materials

| Parameter | Unit | Au [24, 27] | | Ti [18, 19] | Cr [28] | Glass[29, 30] |
|---|---|---|---|---|---|---|
| Density, $\rho$ | kg/m$^3$ | 6950 | | 4500 | 7140 | 2630 |
| Lattice specific heat, $c_p$ | J/(kg·K) | $110 + 0.128 \times T_l - 3.4 \times 10^{-4} \times T_l^2$ $+5.24 \times 10^{-7} \times T_l^3 - 3.93 \times 10^{-10}$ (solid) $\times T_l^4 + 1.17 \times 10^{-13} \times T_l^5$ <br><br> 157.2 (liquid) | | 540 | 460 | 705 |
| Electron specific heat coefficient, $\gamma$ | J/(m$^3$·K$^2$) | 60 | | 319.8 | 194 | - |
| Electron-phonon coupling coefficient at room temperature, $G_0$ | W/(m$^3$·K) | $2.2 \times 10^{16}$ (solid) $2.6 \times 10^{16}$ (liquid) | | $2.6 \times 10^{18}$ | $4.2 \times 10^{17}$ | - |
| Latent heat of melting, $L_m$ | J/kg | $6.37 \times 10^4$ | | $3.37 \times 10^5$ | $4.04 \times 10^6$ | - |
| Melting temperature, $T_m$ | K | 1338 | | 1943 | 2180 | 1100 (transition, $T_g$) |
| Latent heat of vaporization, $L_v$ | J/kg | $1.70 \times 10^6$ | | $9.18 \times 10^6$ | $6.53 \times 10^7$ | - |
| Boiling temperature, $T_v$ | K | 3127 | | 3560 | 2944 | - |
| Lattice thermal conductivity, $k_l$ | W/(m·K) | $321 - 0.0111 \times T_l - 2.75 \times 10^{-5} \times T_l^2$ $-4.05 \times 10^{-9} \times T_l^3$ (solid) $37.7 + 0.0711 \times T_l - 1.72 \times 10^{-5} \times T_l^2$ $-1.06 \times 10^{-9} \times T_l^3$ (liquid) | | 22 | 94 | 1.4 |
| Electron thermal conductivity, $k_e$ | W/(m·K) | Eq. (A6) | | 22 | 94 | - |
| Absorption coefficient, $\alpha$ | m$^{-1}$ | $8.2 \times 10^7$ | | - | - | - |
| Electronic velocity, $v_e$ | m/s | $1.4 \times 10^6$ | | $1.8 \times 10^6$ | $1.6 \times 10^6$ | - |

**Acknowledgments**

BK and HKN appreciate the KAIST analysis center for research advancement (KARA) and the Korea Institute of Machinery and Materials (KIMM) for their supports in experimental measurement and data analysis.

**Funding**

This work was supported by the National Research Foundation of the Republic of Korea (NRF-2012R1A3A1050386); Grant-in-Aid for Scientific Research of Japan (18H01379); and Amada Foundation of Japan (AF-2018219).

**Declaration of competing interest**

The authors declare no competing financial interests.

**Data availability**

The data that support the findings of this study are available from the corresponding authors on request.

**Author contributions**

The project was planned and overseen by S.-W.K. and K.F. in collaboration with Y.-J.K. Experiments were performed by B.K., S.W., S.P., and H.K.N. Numerical simulation was conducted by B.K. and S.W. The ultrashort laser system used for experiment was prepared by Y.K. All authors contributed to the manuscript preparation.